\documentclass[11pt]{article}
\usepackage{mathtools}
\usepackage[utf8]{inputenc}
\usepackage[T2A]{fontenc}
\usepackage[ukrainian,english]{babel}
\title{\textbf{Universal ray method for aberration calculation of spectral and polarization selective optical diffraction systems}}
\author{V.M.Vashchenko, 
		Ye.A. Loza,
		Zh.I.Patlashenko \\ \\
		\texttt{State ecological academy}\\
		\texttt{of post graduate education and management}\\
		\texttt{Ukraine, Kyiv, V.Lypkivskogo str, 35}\\
		\texttt{Loza@bmyr.kiev.ua}}
\date{}
\begin{document}

\maketitle

\begin{abstract}
This paper is dedicated to method for spectral and polarization selective optical systems properties investigations for accurate analysis of nonclassical diffraction gratings properties: aspherical surfaces, variable grating step, variable line shape, etc. Using the proposed method and algorithm an accurate spectral image and apparatus function of Seya-Namioka diffraction monochromator were calculated. Another advantage of the method is its applicability to spectropolarimeters schema analysis.

\end{abstract}

\section{Introduction}

This paper presents an accurate geometrical method for diffraction grating calculation for spectrometers and spectropolarimeters. Classic aberration theories enable calculation of basic schema parameters and minimize their aberrations of 3rd, and, in some cases, of up to 5th order. However, these theories are unable to minimize higher-order aberrations and describe polarization effects in detail \cite{1,2}.

Using the grooved diffraction gratings almost any grating step and line shape may be obtained at aspherical surfaces. Holographic gratings yield significantly lesser quantity of instrumentally scattered light due to absence of unused spectral orders and better quality surface shape. \cite{3}. However in case of a complex interference image on the grating surface their aberration calculation cannot be accurately preformed by classic algorithms that hinders creation of new optical elements.

Such complex optical elements and diffraction gratings are sometimes called nonclassical. They may have aspherical surfaces with variable grating step and/or complex line shape. Fractal gratings \cite{4,5,6}, adaptive optics elements and non-stationary gratings, e.g. electrooptical gratings, may also be considered as nonclassical \cite{7}.

Small-sized spectrometers may achieve higher resolution by using nonclassical optical systems and elements which may significantly reduce optical system aberrations and open new possibilities of spectral image forming, increase the device optical efficiency by high incidence angles. Classic aberration theory does not supply necessary calculation accuracy for such schema bringing up the need for more universal approaches to aberration calculation.

It is known, that there are three main theoretical approaches to diffraction monochromator aberrations calculation.

One of the most accurate approaches is direct Maxwell equations solution. But due to very high method complexity it may be applied only to simple periodic diffraction elements \cite{8,9}.

Application of diffraction optics also enables for spectral image calculation but they do not consider optical properties of material. This creates problems for optical system instrumental polarization calculation. Moreover, accurate calculations for nonclassical elements may also require inappropriate calculation power.

And the third group of methods based on ray optics. Due to simplicity they may be used for calculation of very complex optical elements \cite{10}. Moreover polarization tracking is also possible \cite{3}.

Researchers, engineers and manufacturers require accurate algorithms, that would enable them to determine parameters of monochromators of any complexity level and control those parameters before sample manufacturing. Therefore creation of a new, more universal theory would enable not only calculation accuracy increase and reduce ambiguity of adjusting, but also create a new elementary base and implement new construction solutions.

\section{Mathematical problem set-up}

Let us consider a 3D surface of an aspherical diffraction grating with lines of arbitrary shape. Let's place Cartesian coordinate system center $(0;0;0)$ at its vertex. Let $oz$ axis go along the central line, and $ox$ -- along internal normal to grating surface \cite{11}.

In such geometry XOY plane is called meridional plane and XOZ plane is called sagittal plane \cite{12}. However, unlike in classic theory where these two planes were determined relatively to the whole grating, in this paper we consider them only for reference, because each point at grating surface would have different meridional and sagittal planes depending on local surface and line shape peculiarities.

In order that ray optics may be applied the following condition should be met at each point at the grating surface \cite{13}:

\begin{equation}
\lambda<<e(x,y,z)<<R(x,y,z),
\label{eq1}
\end{equation}

where $\lambda$ -- optical wavelength, $R$ -- grating surface average local curvature radius near point $(x,y,z)$, $e$ -- is distance between two lines at the grating surface near point $(x,y,z)$, i.e. physical grating step. Spectral diffraction order is denoted by $n$. Values $e(x,y,z)$ and $R(x,y,z)$  for nonclassical optical elements are functions of coordinates at the grating surface. In case condition \eqref{eq1} is met the function $e(x,y,z)$ may be considered continuous.

Let every line is a continuous real curve. Let's define a 'line direction' vector $\vec{\tau}$, i.e. tangent to spatial line curve at each point of the grating surface. Let $|\vec{\tau}|=1$ for simplicity. Let's also determine that $\vec{\tau}$ is directed in positive $oz$ axis direction, i.e. $\tau_z>0$. In case of complex line shape an extended condition should include continuity of $\vec{\tau}$ direction along the line, but actually there is no urgent need to consider such case because such gratings would be hard to create, and their benefits are unclear.

\section{Universal 2D ray theory for nonclassical diffraction gratings}

First of all, let us consider a problem of spectral image formation in principal meridional plane of the grating. Mathematically it is the problem of finding intersection of any two rays from a single point at entrance slit which diffract at different elements of the diffraction grating surface (fig.~1). Let one of these rays be 'central', i.e. falling to the grating vertex $(0,0)$. In 2D case this problem has analytical solution \cite{14}:
\begin{equation}
x'_i = \eta_i\zeta_i \qquad \text{and} \qquad y'_i=\zeta_i,
\label{eq2}
\end{equation}
where
\begin{equation}
\zeta_i = y'_i = \frac{y-x\vartheta_i}{1-\eta_i\vartheta_i};
\end{equation}
\begin{equation}
\eta'_i=\sqrt{1/\Big(\frac{n \lambda_i}{e(y)}\cos\theta-\sin \psi_y\Big)^2-1};
\label{eq4}
\end{equation}
\begin{equation}
\eta_i=\eta'_i(y=0);
\end{equation}
\begin{equation}
\vartheta_i = \frac{(R_x(y)-x)-\eta'_i(R_y(y)-y)}{ (R_x(y)-x) \eta'_i + (R_y(y)-y)},
\label{eq5}
\end{equation}
where
\begin{equation}
R_x(y) = x(y) + R(y)\cos\theta;
\end{equation}
\begin{equation}
R_y(y) = y - R(y) \sin\theta;
\end{equation}
\begin{equation}
R(y) =\frac{\bigg [1+\Big(\frac{dx}{dy}\Big)^2\bigg]^{3/2}}{\frac{d^2x}{dy^2}}
\label{eq6}
\end{equation}

where $R(y)$ is surface curvature radius at point y, $\psi_y$ -- incidence angle in the diffraction plane (in our case -- the primary meridional plane) relative to the grating surface normal at the incidence point. The angle is determined by:
\begin{multline}
\sin^2\psi_y = 1 - \frac{R(0)^2 R_x(y)^2}{[(x-R_x(y))^2+(y-R_y(y))^2][(x-R(0))^2+y^2]}
\label{eq7}
\end{multline}

$\varphi$ is diffraction angle and $\theta$ determines the angle of tangent to the grating surface relative to $oy$ axis and is determined by $\tan(\theta)=dx/dy$ \cite{11}.

\begin{figure}
\begin{center}
\includegraphics[width=0.5\textwidth]{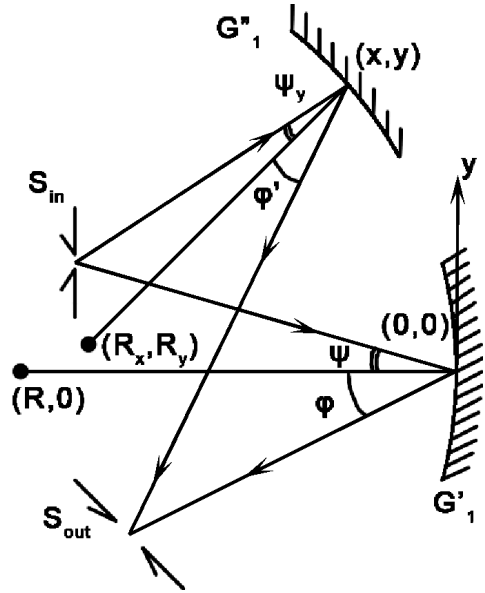}
\vskip-3mm\caption{Ray diffraction at different grating surface elements $\text{G}'_1$ and $\text{G}''_1$. $S_{in}$ and $S_{out}$ are entrance and exit slits, $\phi$ and $\psi$ are incidence and diffraction angles for central ray respectively, and $\phi'$ and $\psi_y$ are the same values for non-central ray.}
\end{center}
\end{figure}

As a result we obtain a set of intersection points for a set of chosen rays pairs that form spectral image. And after obtaining an analytic representation of spectral image we may optimize it. The tasks may be e.g. to achieve 'flat field'. This problem is solved by a non-linear algebraic equation to determine local line density $e(x,y,z)$ \cite{14}:
\begin{equation}
\frac{\eta_2-\eta_3}{\zeta_1}+ \frac{\eta_3-\eta_1}{\zeta_2}+ \frac{\eta_1-\eta_2}{\zeta_3}=0
\label{eq8}
\end{equation}
where indexes 1, 2 and 3 correspond to three different wavelengths of a given spectral range. Numerical solution for \eqref{eq8} is given in fig.~2.

\begin{figure}
\begin{center}
\includegraphics[width=0.7\textwidth]{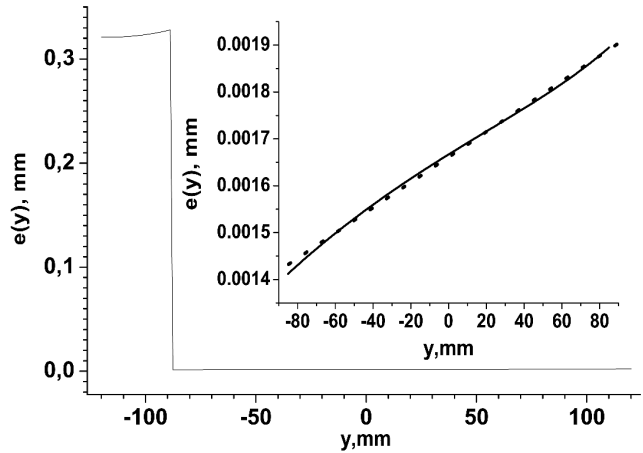}
\vskip-3mm\caption{Graphical solution of eq. \eqref{eq8}. Classical aberration theory provides for linear grating step change (dotted line).}
\end{center}
\end{figure}

The result of best focusing 'volume' optimization for spherical diffraction grating with variable grating step is represented at fig.~3. By best focusing volume we mean a volume of best spectral image sharpness, i.e. a set of intersection points for rays pairs forming spectral image.

A more detailed analysis of best focusing volume for a classic grating shows that at intersection of spectral orders the volume shape is equal. For nonclassical grating best focusing volumes are different for different spectral orders and have an additional bend in all but 1st order. At small wavelengths they intersect at point corresponding to mirror image that would have been created by a spherical mirror. There is a 'limit' wavelength $\lambda=e_0/n \cdot (1+\sin\psi_y)$ for which diffraction angle is 90$^o$.

Astigmatic difference increases linearly with wavelength in the optimization spectral region. Beyond the optimization spectral region there are two bends for best focusing volume and astigmatic difference increases near those.

Such 'optimized' nonclassical grating with variable grating step may be used only in one spectral order it was calculated for. It should also be noted that astigmatic difference increases by +30\% to +10000\% for such grating \cite{14}.

Variable grating step may also be used for aberration correction. This problem renders a differential equation:
\begin{equation}
d\zeta/dy = 0,
\label{eq9}
\end{equation}
which can be rewritten as \cite{14}:
\begin{multline}
\frac{de(y)}{dy} = \bigg(\Big[
\big(1-\vartheta\frac{y}{R-x}\big)\frac{1-\eta\vartheta}{\eta y - x} \frac{((R-x)\eta'-y)^2}{R^2}+\\+\frac{\eta'^2-1}{R-x}\Big]\cdot\eta'\big(\frac{n\lambda}{e(y)}\cos\theta-\sin\psi_y\big)^3 -\frac{d\sin\psi_y}{dy}\bigg) \frac{e(y)^2}{n\lambda} - \\ - e(y)\frac{y/R^2}{\cos\theta}.
\label{eq10}
\end{multline}
The equation \eqref{eq10} can be solved by numerical methods \cite{15}.

The complete aberration correction in meridional plane may be achieved by diffraction grating surface shape. This problem is formalized as implicit differential equation of third order. However, such diffraction grating shape would hardly be technologically rational.

\section{Universal 3D ray theory for nonclassical diffraction gratings}

The next step of aberration analysis is a 3D geometrical theory enabling spectral image calculation to investigate and optimize their characteristics. The core problem is transformation of incident ray to a diffracted ray considering local surface peculiarities (fig.~4).

In the chosen coordinate system the grating surface is determined by equation:
\begin{equation}
F(x,y,z) = 0,
\label{eq11}
\end{equation}

\begin{figure}
\begin{center}
\includegraphics[width=0.5\textwidth]{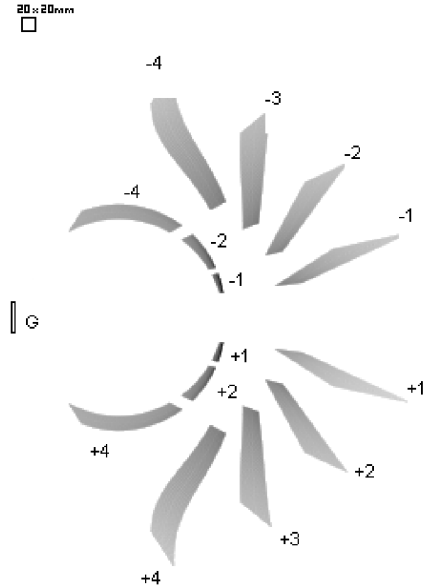}
\vskip-3mm\caption{Spatial distribution of plane-like volumes of best focus at different spectral orders for classical and nonclassical spherical diffraction gratings (G) in spectral range 200 to 350 nm. A square above the figure shows the scale.}
\end{center}
\end{figure}

\begin{figure}
\begin{center}
\includegraphics[width=0.7\textwidth]{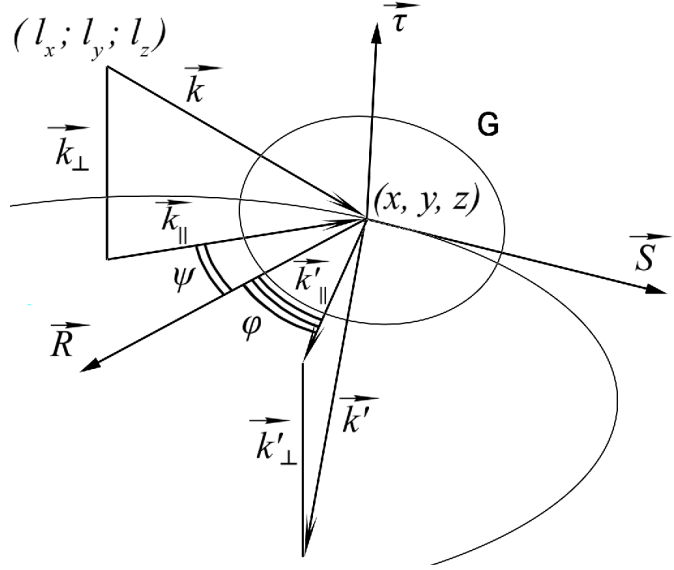}
\vskip-3mm\caption{Ray diffraction at grating G. There is no need to specify entrance and exit slits.}
\end{center}
\end{figure}

Central concept of the 3D theory is diffraction plane \cite{12}. Let us subdivide the grating surface into a large number of small quasi-spherical classic diffraction gratings each having its size significantly less than its curvature radius $R$. For small working surface size classic diffraction grating formula would provide high accuracy.

As a result, we may define diffraction plane as a plane that passes through ray incidence point at the grating surface and is perpendicular to the line direction $\vec{\tau}$ at this point \cite{16}.

Intersection if diffraction plane and the grating surface forms a spatial curve. Its curvature radius $R$ is equal to Rowland circle diameter \cite{17}. Therefore as a result we obtain multiple Rowland circles with different diameters $R$ and spatial orientation for each point at diffraction grating surface, and a corresponding set of meridional and sagittal planes. Local Rowland circle diameter would be equal to \cite{13}:
\begin{equation}
-\vec{R} = [\vec{S}\times\vec{\tau}]\cdot|\vec{S}|^2/\vec{\tau}\cdot \vec{B}
\label{eq12}
\end{equation}
where $\vec{S}$ and $\vec{B}$ -- tangent and binormal respectively \cite{11}.

Change of an incident ray $\vec{k}$ into a diffracted ray $\vec{k’}$ is described by generalized diffraction grating equation \cite{10}
\begin{equation}
(\vec{k}_\parallel)_{\vec{S}} + (\vec{k}'_\parallel)_{\vec{S}} = n\lambda / e(x,y,z)
\label{eq13}
\end{equation}
where $\vec{S}$ index denotes projection to tangent vector and $\vec{k}’_\parallel=\hat{C}^{\vec{\tau}}_{\varphi-\psi}\vec{k}_\parallel$, where $\hat{C}^{\vec{\tau}}_{\varphi-\psi}$ -- mathematical rotation matrix by angle $\varphi-\psi$ considering the rule of signs \cite{18} around $\vec{\tau}$, and $\varphi$ -- diffraction angle. This enables us to calculate meridional plane diffracted ray vector component. Sagittal plane component follows mirror reflection $\vec{k}_\perp=\vec{k}'_\perp$. And finally:
\begin{equation}
\vec{k}'=\vec{k}'_\parallel+\vec{k}'_\perp
\label{eq14}
\end{equation}

In order to operate the obtained model a numerical representation of diffraction grating surface should be introduced to conveniently change local and global diffraction grating parameters (such as grating surface local curvature, local grating step and line shape) in optimization tasks.

Let us introduce a set of points $(y_{i,j};z_{i,j})$ to describe diffraction grating surface defined by \eqref{eq11} (fig.~5) which determines $x_{i,j}=x(y_{i,j};z_{i,j})$. Then the set of $(x_{i,j},y_{i,j};z_{i,j})$ determine points at the grating surface. Zero approximation is determined by simplified approach, e.g. by 2D grating calculation. Let the number of lines between each node along  $oy$ axis is constant and equals to $N$.

\begin{figure}
\begin{center}
\includegraphics[width=0.5\textwidth]{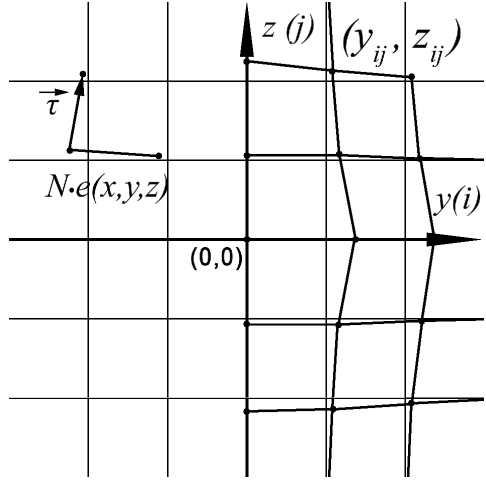}
\vskip-3mm\caption{Set of points at the grating surface for numerical model.}
\end{center}
\end{figure}

The line direction would be determined by:
\begin{equation}
\vec{a}=\{x_{i,j+1}-x_{i,j-1}; y_{i,j+1}-y_{i,j-1}; z_{i,j+1}-z_{i,j-1}\}
\label{eq15}
\end{equation}
\begin{equation}
\vec{\tau}_{i,j} = \vec{a} / |\vec{a}|.
\end{equation}
The proposed theory enables calculation (fig.~6) and optimization (fig. ~7) of the spectral image.

\begin{figure}
\begin{center}
\includegraphics[width=0.8\textwidth]{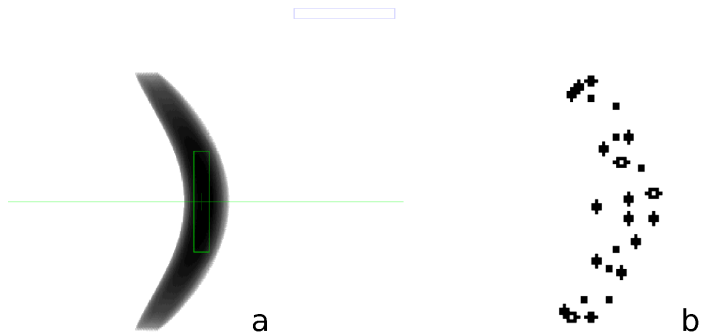}
\vskip-3mm\caption{Spectral image: a -- calculated by the described algorithm for 4th spectral order of spherical diffraction grating with 600 lines per mm at 300 nm wavelength; b -- calculated for similar setting by classical raytracing approach \cite{3}.}
\end{center}
\end{figure}

\begin{figure}
\begin{center}
\includegraphics[width=0.8\textwidth]{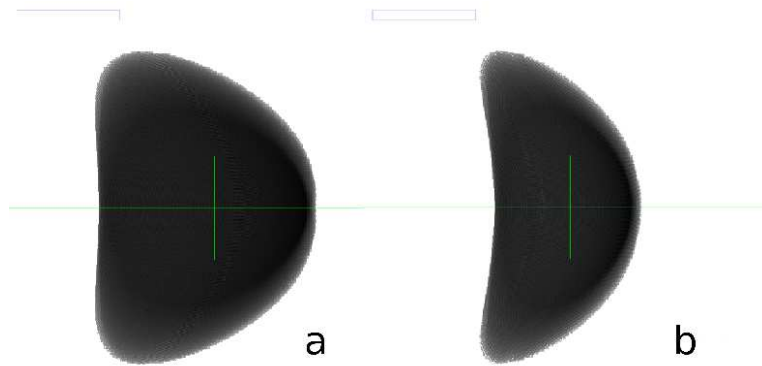}
\vskip-3mm\caption{Spectral image calculated for 600 lines per mm spherical grating at 350 nm wavelength before (a) and after (b) focal plane location optimization.}
\end{center}
\end{figure}

Moreover, based on calculation results, the apparatus function of the monochromator may be plotted and optimized (fig.~8).

In order to reduce aberrations classical theory proposes application of elliptical apertures at grating surface. However, our result show that, while boosting image quality by around 2\% such approach reduces grating optical efficiency by 12\%. Therefore, despite visually more 'beautiful' spectral image, such technological approach is not efficient. Image calculation in case of curved entrance slits classically used to correct spectral image curvature also appears to be inefficient, due to significant apparatus function width increase while correcting the image curvature.

\begin{figure}
\begin{center}
\includegraphics[width=0.6\textwidth]{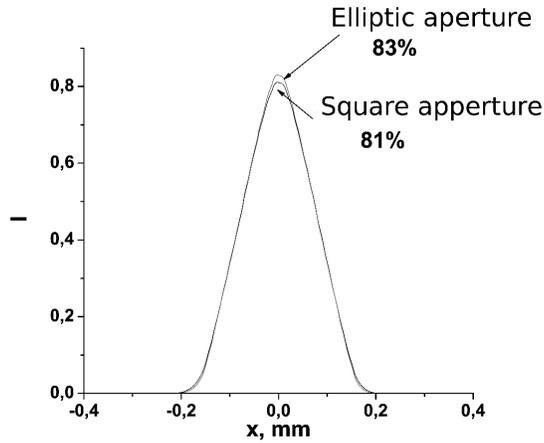}
\vskip-3mm\caption{Comparison of calculated normalized apparatus functions for square and elliptic apertures at the grating surface for 1st spectral order.}
\end{center}
\end{figure}

Our approach enables apparatus function assymetry calculation. E.g. fig.~9 shows apparatus function of spectral image at fig.~6.

\begin{figure}
\begin{center}
\includegraphics[width=0.6\textwidth]{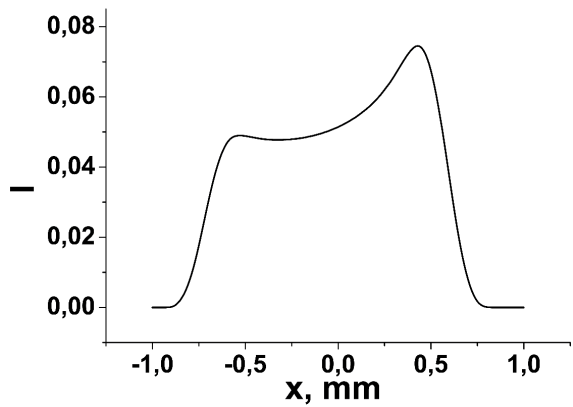}
\vskip-3mm\caption{Apparatus function asymmetry for 4th spectral order of the 600 lines per mm spherical grating at 300 nm wavelength.}
\end{center}
\end{figure}

The presented 3D aberration theory in contrast to classical approach to aberration calculation is applicable for large grating angular size, i.e. for large incidence angles. It may be successfully applied to variable grating step, non-linear line shape, complex aspherical surfaces and adaptive optics elements. The theory and algorithm enables calculation and optimization of spectral images and apparatus functions at calculation stage.

\section{Polarization calculation peculiarities}

Modern spectrometry is limited to one Stokes parameter (i.e. intensity) and therefore cannot be used to solve many new physical problems. Therefore modern trend in atmospheric investigations is spectropolarimetry, enabling 2 or more Stokes parameters measurements at different wavelengths \cite{21}.

In order to calculate a spectropolarimeter schema an accurate polarization impact of the monochromator must be known. Neglecting the instrumental polarization can cause errors estimated to be up to 70\% \cite{22} which is sometimes solved by neutralizing polarization impact by symmetrization of the double monochromators and application of additional polarization and depolarization elements \cite{22}.

However, the described algorithms enable not only calculation of instrumental polarization inside the monochromators, but also control it at theoretical stage. It enables deliberate shaping of instrumental polarization to fit some pre-determined state and therefore reduce the number of required optical elements for polarization measurements implementation.

In order to calculate polarization influence of the monochromator, each ray is attributed with Stokes vector and a theoretically calculated or experimentally measured Muller matrix is attached to each optical surface element in the monochromator.

As a result instrumental polarization may be calculated for each wavelength. Its distribution along the spectral image may also be plotted to be used in optimization procedure and inverse problem solution.

Moreover, the optical system may imply adaptive polarization impact at different wavelengths and spectral image parts by magnetic or electrooptical elements.

\section{Conclusions}

1. The proposed new universal mathematically rigid ray aberration theory for nonclassical focusing diffraction gratings is applicable within ray optics validity conditions. It may be used for analysis of nonclassical optical elements with aspherical surfaces, variable grating step, complex line shape, adaptive optics elements, high incidence angles, etc.

2. The developed algorithm provides for accurate calculation and optimization of spectral image during scheme theoretical calculation stage in accordance to technical requirements determined by the physical problems to be solved.

3. The engineer and physics problems of optical aberrations minimization and efficiency are formulated as implicit algebraic and differential equations and enable numerical calculation of 2D and 3D spectral images and also 1D and 2D apparatus function calculation and optimization.

4. The proposed numerical model for nonclassical diffraction grating enables mathematical control of the grating surface for convenient optimization of target function and design principally new optical elements and schema.

5. The algorithm is applicable for high-quality polarization calculation of monochromator schema and creation of principally new spectropolarimeters for measuring 2 to 4 Stokes parameters.


\begin{thebibliography}{22}

\bibitem{1}Shin Masui and Takeshi Namioka
JOSA A, Vol. 16, Issue 9, pp. 2253-2268 doi:10.1364/JOSAA.16.002253
\bibitem{2}J. F. Seely, R. G. Cruddace, M. P. Kowalski, W. R. Hunter, T. W. Barbee, Jr., J. C. Rife, R. Eby, and K. G. Stolt
Applied Optics, Vol. 34, Issue 31, pp. 7347-7354 doi:10.1364/AO.34.007347
\bibitem{3}Christopher Palmer, Erwin Loewen 
Thermo RGL; sixth edition, 2005, 265P
\bibitem{4}Dongsu Bak, Sang Pyo Kim, Sung Ku Kim, Kwang-Sup Soh, Jae Hyung Yee
arXiv:physics/9802007
\bibitem{5}D. Rodriguez Merlo, J. A. Rodrigo Martin-Romo, T. Alieva  and M. L. Calvo
Optics and Spectroscopy, Volume 95, Number 1 / July, 2003, P.131-133, DOI 10.1134/1.1595227
\bibitem{6}C. Aguirre Veleza, M. Lehmana and M. Garavaglia
Optik - International Journal for Light and Electron Optics, Volume 112, Issue 5, 2001, Pages 209-217, doi:10.1078/0030-4026-00038
\bibitem{7}J. Chen, P. J. Bos, H. Vithana, and D. L. Johnson
Appl. Phys. Lett. 67, 2588 (1995); doi:10.1063/1.115140
\bibitem{8}LIN Albert, PHILLIPS Jamie
Solar energy materials and solar cells   ISSN 0927-0248 2008, vol. 92, no12, pp. 1689-1696
\bibitem{9}Eero Noponen
Dissertation for the degree of Doctor of Technology to be presented with due permission for public examination and debate in Auditorium F1 at Helsinki University of Technology (Espoo, Finland) on the 15th of April, 1994, at 12 o’clock noon. Espoo 1994
\bibitem{10}Ващенко В.М., Лоза Є.А., Патлашенко Ж.І., Банніков О.О., Черниш О.Є.
Вісник Київського університету, серія: фізико-математичні науки. - 2005 - №4 - С.430-440.
\bibitem{11}Бронштейн Н.Н., Семендяев К.А.
Государственное издательство технико-теоретической литературы, Москва, 1953г - 608с.
\bibitem{12}Зайдель А.Н., Островская Г.В., Островский Ю.И.
Москва, Наука, 1976, 375с.
\bibitem{13}М.Борн, Э.Вольф
М:Наука, 1973, 721с.
\bibitem{14}Ващенко В.М., Лоза Є.А., Патлашенко Ж.І.
Вісник Київського університету, серія: фізико-математичні науки. - 2008 - №4 - С.245-251.
\bibitem{15}Калиткин Н.Н.
Москва, Наука, 1978, 508с.
\bibitem{16}Ващенко В.М., Лоза Є.А., Патлашенко Ж.І.
Вісник Київського університету, серія: фізико-математичні науки. - 2009 - №2 - С.235-242.
\bibitem{17}Корн Г., Корн Т.
Издательство «Наука», Москва, 1973, 831 с.
\bibitem{18}Одарич В.А.
Київ, видавничо-поліграфічний центр Київський університет, 2001, 220 с.
\bibitem{21}Vaschenko V., Patlashenko Zh., Chernysh E.
Semiconductor physics. Quantum electronics and optoelectronics - 2004 - vol.7 №1 - P.105-107.
\bibitem{22}Vaschenko V.N., Loza A.I., Patlashenko J.I.
Proceedings of SPIE - 1997 - vol. 3237 - P.31-42.
\end{thebibliography}
\end{document}